\documentclass[a4paper]{article}

\usepackage{INTERSPEECH2020}
 \usepackage{booktabs}
 \usepackage{multirow}
 \usepackage{multicol}
\title{Transfer Learning Approaches for Streaming End-to-End Speech Recognition System}
\name{Vikas Joshi, Rui Zhao, Rupesh R. Mehta, Kshitiz Kumar, Jinyu Li}
\address{Microsoft Corporation}
\email{vikas.joshi, ruzhao, rupesh.mehta, kshitiz.kumar, jinyli@microsoft.com}

\begin{document}

\maketitle
\begin{abstract}

Transfer learning (TL) is widely used in conventional hybrid automatic speech recognition (ASR) system, to transfer the knowledge from source to target language. TL can be applied to end-to-end (E2E) ASR system such as recurrent neural network transducer (RNN-T) models, by initializing the encoder and/or prediction network of the target language with the pre-trained models from source language. In the hybrid ASR system, transfer learning is typically done by initializing the target language acoustic model (AM) with source language AM. Several transfer learning strategies exist in the case of the RNN-T framework, depending upon the choice of the initialization model for encoder and prediction networks. This paper presents a comparative study of four different TL methods for RNN-T framework. We show ~$10\%-17\%$ relative word error rate reduction with different TL methods over randomly initialized RNN-T model. We also study the impact of TL with varying amount of training data ranging from $50$ hours to $1000$ hours and show the efficacy of TL for languages with a very small amount of training data.\\

\end{abstract}

\section{Introduction}
Speech enabled applications are increasingly gaining popularity across the world. This has initiated a need to build accurate automatic speech recognition (ASR) system across different languages. Also, End-to-End (E2E) ASR systems are emerging as a popular alternative to conventional hybrid ASR systems. They replace the acoustic model (AM), language model (LM) and pronunciation model with a single neural network \cite{RNNT_graves,Graves_CTC_init,LAS,E2E_comp_1,E2E_comp_2,E2E_comp_3,E2E_comp_4,E2E_comp_7,Li2019RNNT, sainath2020streaming, Li2020Comparison}. Recurrent neural network transducer (RNN-T) \cite{RNNT_graves} is one such E2E system that allow streaming input and is suitable for real-time ASR applications. Therefore there is a lot of interest in building accurate RNN-T models for different languages spoken across the world.

%The speech recognition accuracy largely depends upon the amount of training data available. 
There is often disparity in the availability of transcribed data for different languages. In most cases, a lot more data is available for American English than other languages. The quality of ASR model depends on a number of factors including, the training data quantity and diversity, acoustic model structure, and  optimization algorithm. Furthermore, training data diversity spans a number of factors in adults,  kids, speaking rate,  accents, near-field, and far-field acoustic conditions. A low-resource locale has limited ASR training data, and may not meet the acoustic diversity needed to train a robust model that can generalize to above acoustic factors. To overcome the low-resource constraint, transfer learning has been widely used in the hybrid ASR system to transfer the knowledge from a well trained source locale to a low-resource target locale that bring significant acoustic robustness for the target locale. In our recent work, we applied TL from a large scale en-US conventional hybrid model to the corresponding models in en-IN and it-IT locales, and achieved over 8\% word error rate relative reduction (WERR). Motivated by the success of the TL methods in the hybrid ASR system, we explore TL methods to improve low-resource RNN-T models.

Besides improving the target model acoustic robustness, TL is also crucial for training large and complex deep learning architectures. RNN-T models are difficult to train \cite{hu_ce_init} and also require significantly large amount of data to jointly train the acoustic as well as language model attributes. In our study we have noted weaker convergence or significant parameter tuning requirements for desirable E2E training outcome for low-resource locale. Therefore we expect TL techniques to be even more relevant for E2E systems to stabilize training and improve ASR accuracy.

In the hybrid ASR system, transfer learning is typically done by initializing the target AM with the source AM. In the RNN-T framework, several transfer learning strategies exist depending upon the choice of the initialization model for the encoder and prediction networks. In this paper, we compare different transfer learning strategies in the RNN-T framework. We propose two-stage TL, by first training a target initialization model bootstrapped with a pretrained source model. Subsequently, this model is used to initialize the target RNN-T model. The two-stage TL approach shows ~$17\%$ WERR reduction and faster convergence in the training loss as compared to randomly initialized RNN-T model. We also study the effect of TL with different amount of training data and show the importance of transfer learning in the case of low-resource languages.

\section{Relation to prior work}
Several methods have been proposed to improve the performance of low-resource ASR models  \cite{MLT_1,swietojanski2012unsupervised, MLT_2,MLT_3,TL_1,TL_2,SHL_1,SHL_2,EL_1,EL_2}. Successful strategies include transfer learning  \cite{TL_1,TL_2}, that leverage a well trained AM from high-resource language to bootstrap the low-resource AM; multi-task training \cite{SHL_1,SHL_2} and ensemble learning \cite{EL_1, EL_2} that aim to utilize multi-lingual data and share the model parameters. However, most of these methods are studied in the context of hybrid ASR system.

A few multi-lingual approaches are recently proposed in the E2E framework \cite{Multilingual_RNNT,bytes_E2E,MLT_RNNT_Umesh}.  Authors in  \cite{Multilingual_RNNT} propose the multi-lingual RNN-T model with language specific adapters and data-sampling to handle data imbalance. Audio-to-byte E2E system is proposed in \cite{bytes_E2E} where bytes are used as target units instead of grapheme or word piece units, as bytes are suitable to scale to multiple languages. A transformer based multi-lingual E2E model, along with methods to incorporate language information is proposed in \cite{MLT_RNNT_Umesh}. Although multi-lingual methods are attractive to address the problem of low-resource languages, the transfer learning methods, besides being simple and effective, have the benefit of not needing the high-resource language data, but only the models trained on them. In many practical scenarios, trained models are available, however the original corpus is not. Given the simplicity and effectiveness of TL, we explore transfer learning approaches to improve the performance of low-resource RNN-T models.

The rest of this paper is organized as follows: In Section \ref{rnnt_arch}, we briefly
discuss the RNN-T model. The transfer learning methods for RNN-T are described in Section \ref{tl_rnnt} and experimental setup in Section \ref{expt_setup}. Next, we discuss results in Section \ref{discussion_res},
followed by conclusions in Section \ref{conclusion}.

\section{RNN Transducer model}
\label{rnnt_arch}

\begin{figure}
%\begin{center}
\hspace{-0.9cm}
\includegraphics[scale=0.46]{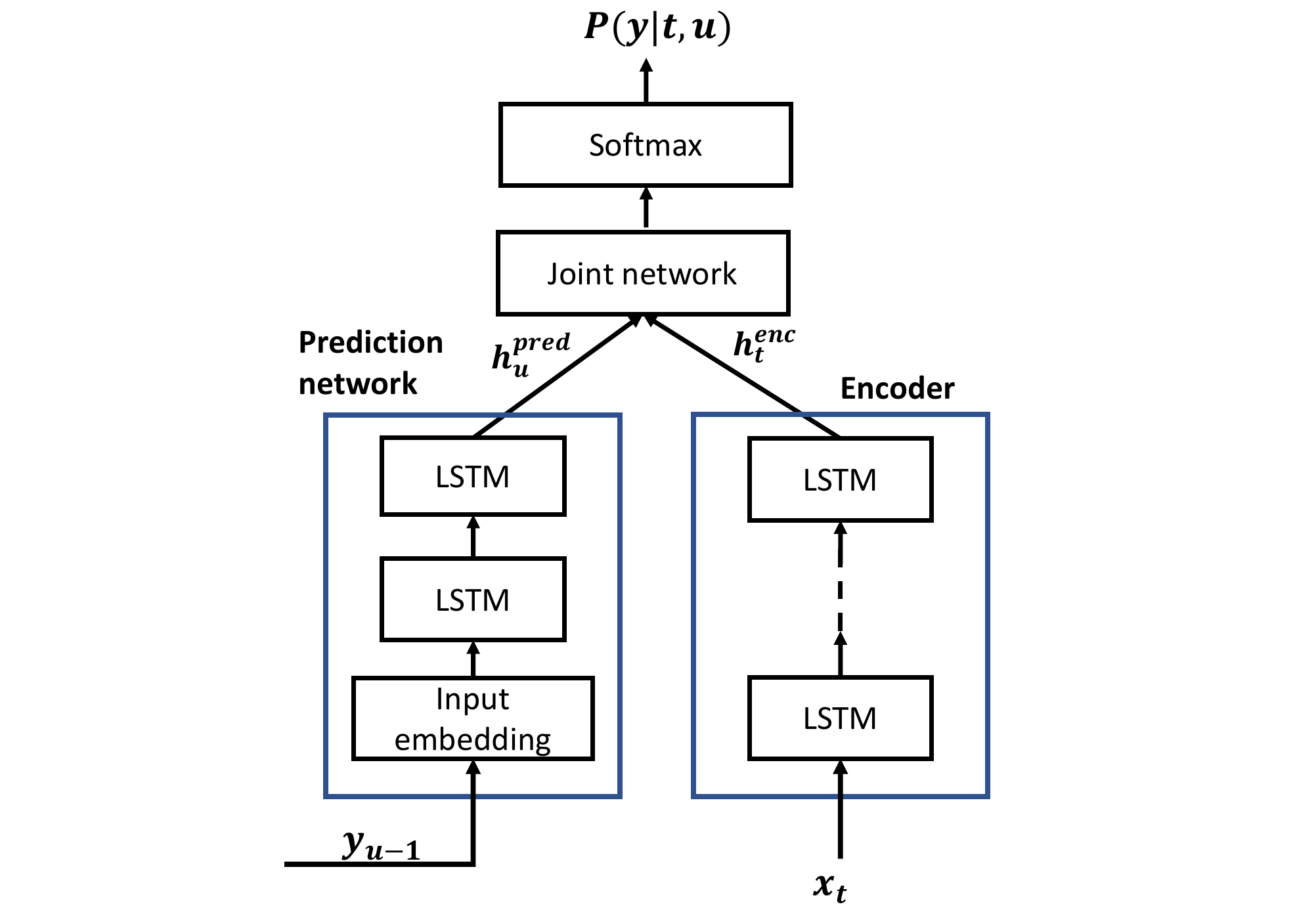}
%\end{center}
\vspace{-0.5cm}
\caption{The RNN-T model.}
\label{rnnt_arch_fig}
\vspace{-0.5cm}
\end{figure}

The RNN-T model was proposed by Alex Graves \cite{RNNT_graves}. The RNN-T model architecture has three components; an encoder, prediction network and joint network as shown in Fig. \ref{rnnt_arch_fig}. The encoder maps the input acoustic feature, $x_t$, to a high level representation, $h{_t}^{enc}$, where $t$ represents the time index. The prediction network receives the previously predicted non-blank symbol, $y_{u-1}$ and maps it to a high level representation, $h_{u}^{pred}$. The joint network is a feed forward network that combines the encoder and  prediction network outputs. The posterior probability over all the targets, $p(y|t,u)$ is obtained after softmax operation on the output of joint network. The whole network is trained jointly to minimize the RNN-T loss \cite{RNNT_graves}. In our implementation, the encoder consists of $6$ long short-term memory (LSTM) \cite{Hochreiter1997long}  layers and the prediction network has $2$ LSTM layers along with the input embedding matrix.  During inference, beam search decoding is used to find the most likely label sequence.

\section{Transfer learning methods for RNN-T}
\label{tl_rnnt}
The RNN-T models are difficult to train and are often initialized with the pretrained models. Initializing the encoder with  connectionist temporal classification (CTC) model \cite{Graves_CTC_init} or cross entropy (CE) model \cite{hu_ce_init}, and the prediction network with LSTM language model (LM) is proven to be beneficial \cite{Rao_hierarchical}. Transfer learning can also be used to overcome the RNN-T training difficulty by initializing the low-resource (target) RNN-T models with the models trained on high-resource (source) languages.

A number of choices exist in selecting the initialization model for encoder and prediction network in the context of TL the RNN-T model. Authors in \cite{hu_ce_init} have shown that CE initialized RNN-T models perform better than CTC initialized models, and hence, we only explore CE models for initialization. The following choices exist for encoder/prediction network initialization of the target RNN-T model: a) Source RNN-T encoder/prediction networks b) Pretrained networks used to initialize the source RNN-T model c) Pretrained models trained only on the target language. Therefore several combinations are possible depending upon the choice of the initialization model for encoder and prediction network.

In this paper, we explore TL methods in the context of Hindi as the target language and American English as the source language. The goal is to improve Hindi RNN-T model by leveraging models trained on American English, which has approximately ten times more data than Hindi. `en-US' prefix is used to refer to models trained with American English and `hi-IN' prefix is used to refer to models trained with Hindi data. We next discuss different transfer learning strategies in detail. 
%We use grapheme units as target tokens to train the hi-IN RNN-T model. The en-US model is trained with word piece targets as discussed in \cite{hu_ce_init}.

\begin{figure}
%\begin{center}
\hspace{-0.5 cm}
\includegraphics[scale=0.45]{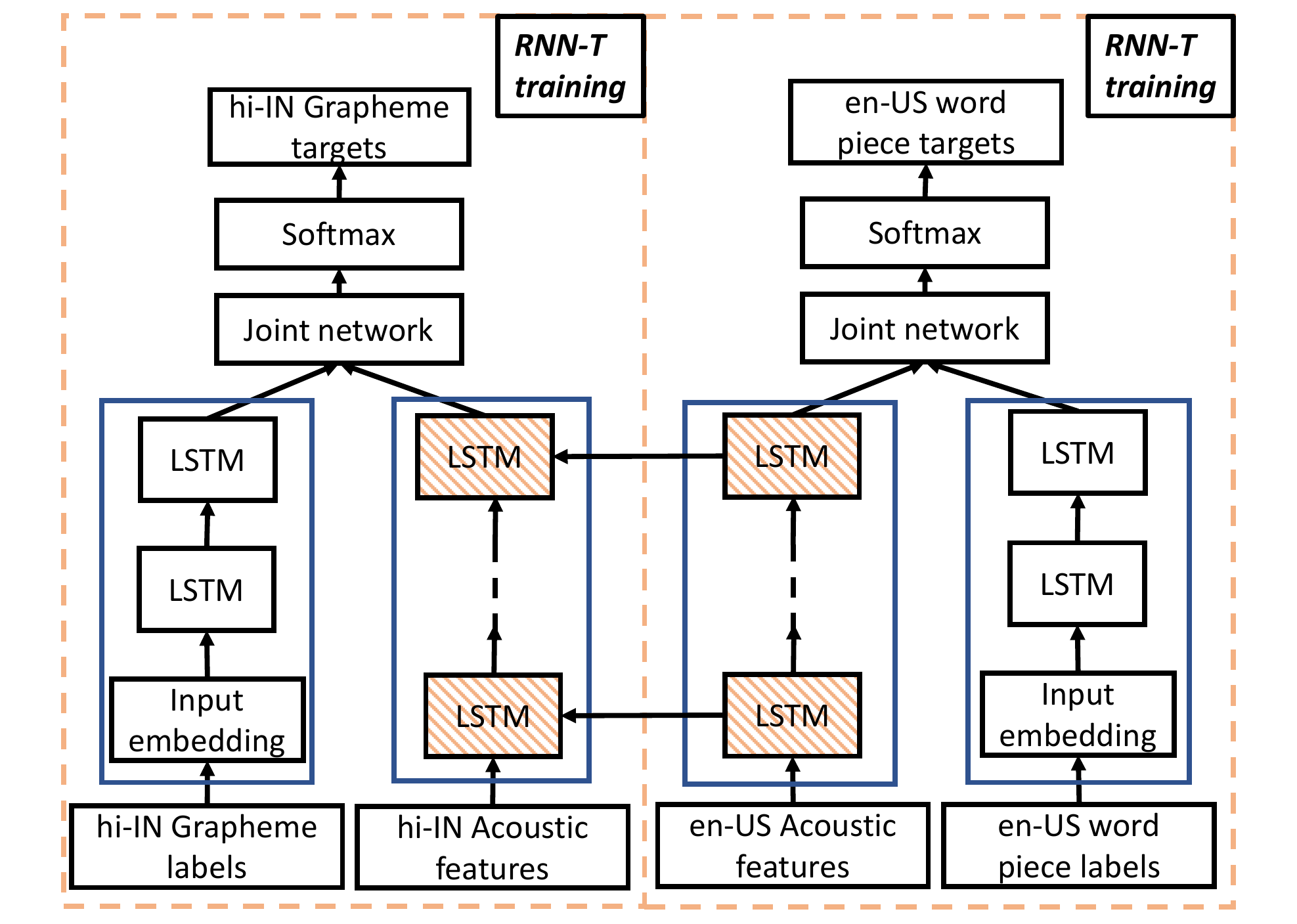}
%\end{center}
\caption{en-US RNN-T initialization.}
\label{enUS_RNNT}
\end{figure}

\subsection{en-US RNN-T initialization}
\label{enUS_rnnt}
The hi-IN RNN-T encoder is initialized with en-US RNN-T encoder as shown in Fig.\ref{enUS_RNNT}. The hi-IN RNN-T model is trained with hi-IN acoustic data and grapheme targets. The en-US RNN-T model is trained with en-US acoustic data and the corresponding word piece targets. The encoder of the en-US RNN-T model is in turn initialized with a pretrained en-US CE model. Note that the prediction network of both hi-IN and en-US RNN-T models are randomly initialized. After initialization, all parameters of the RNN-T model are trained to minimize the RNN-T loss. In all the figures, layers initialized with pretrained networks are represented by cross lined blocks and randomly initialized layers are represented with plain blocks. The details of how we develop en-US RNN-T model can be found in \cite{Li2020Developing}.

\subsection{en-US CE initialization}
\label{enusce_ini}
In en-US CE initialization, the hi-IN encoder is initialized with en-US CE model which was used to initialize the en-US RNN-T encoder, discussed in Section \ref{enUS_rnnt}. The en-US CE model is trained on en-US acoustic data and the corresponding word piece targets. The frame level alignment with word piece targets (necessary for CE training), is obtained from word level alignments as discussed in \cite{hu_ce_init}. From the word alignments, the start frame, end frame and total number of frames corresponding to each word is known. The words are then divided into corresponding word pieces, and equal number of frames are allocated to each word piece within the boundary of the frames corresponding to the word. In this scheme, the hi-IN RNN-T prediction network is randomly initialized. 

\subsection{Two-stage transfer learning}
\begin{figure}
%\begin{center}
\hspace{-0.5cm}
\includegraphics[scale=0.44]{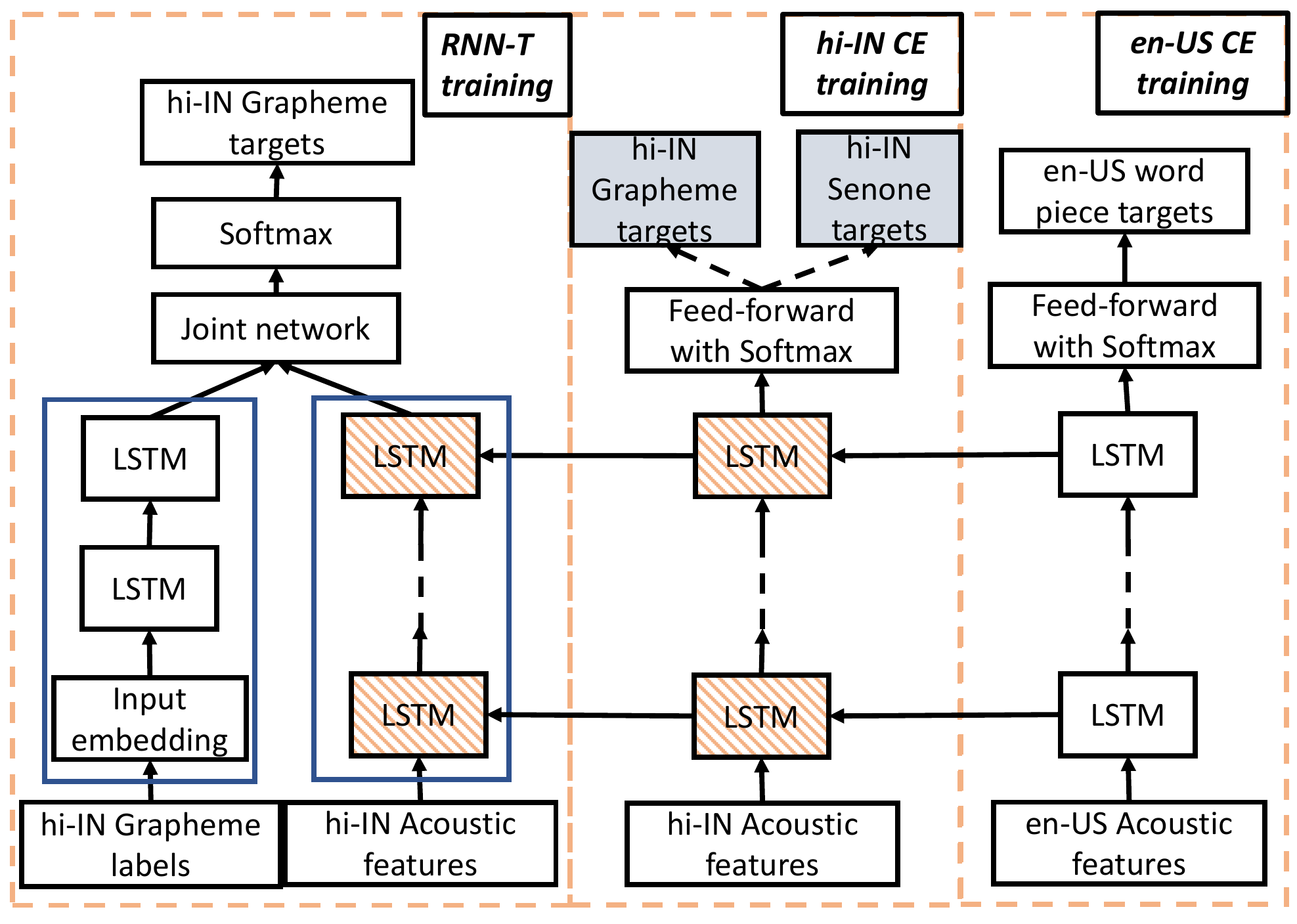}
%\end{center}
\caption{Two stage transfer learning. }
\label{TwoStageTL}
\end{figure}

 Transfer learning can be done in two stages as shown in Fig. \ref{TwoStageTL}. In the first stage, hi-IN CE model is trained starting from en-US CE model. Subsequently, the hi-IN RNN-T model is trained by initializing the encoder with hi-IN CE model. The hi-IN CE model can be trained either with senone and grapheme targets as depicted in Fig. \ref{TwoStageTL}. The senone based CE model can distinguish more finer acoustic classes as senones represent much finer acoustic information than graphemes. However, the grapheme based CE model is better aligned with the RNN-T model as they both are trained with grapheme targets. The prediction network is again randomly initialized.

\begin{figure}
%\begin{center}
\hspace{-0.5cm}
\includegraphics[scale=0.43]{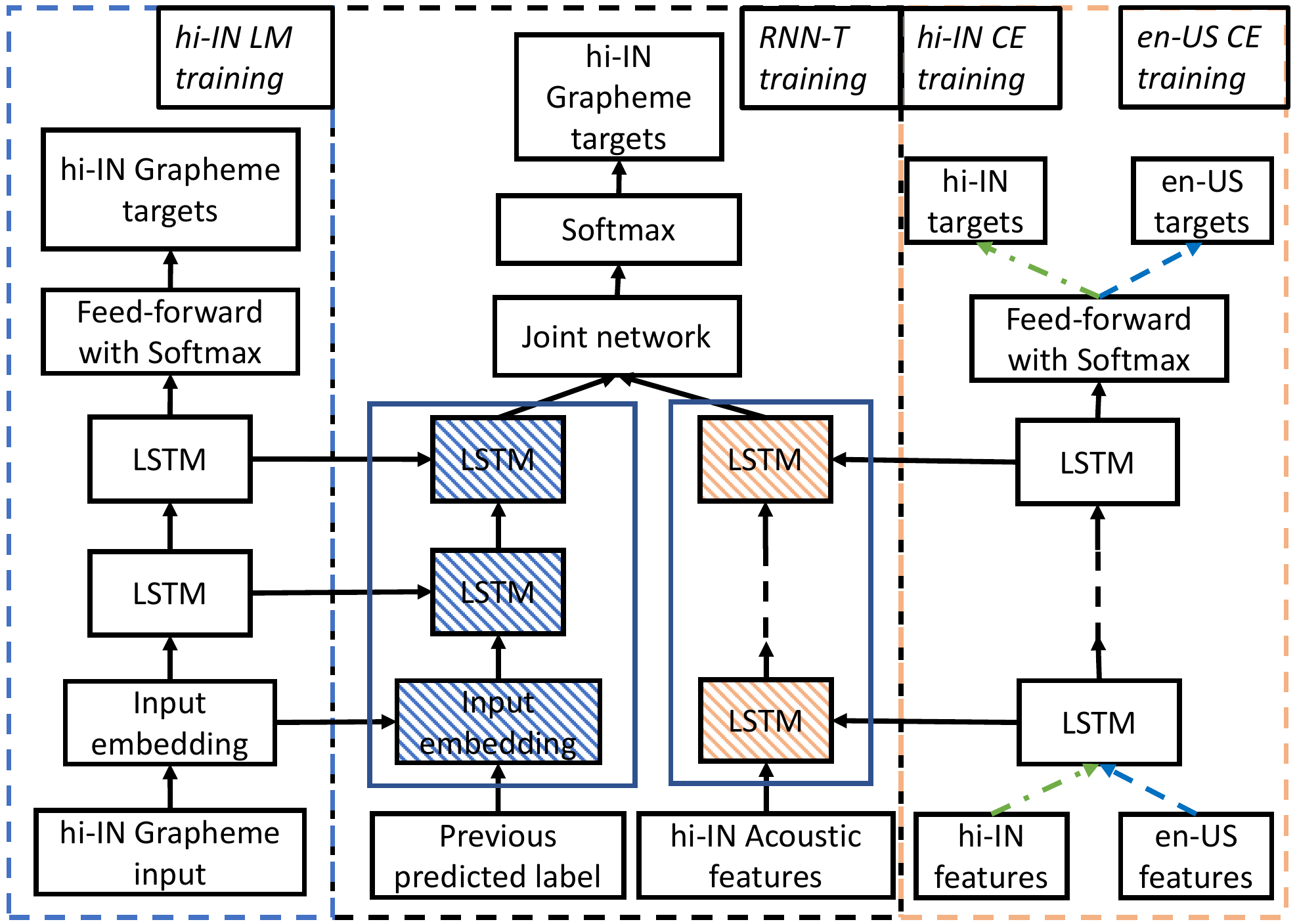}
%\end{center}
\caption{Encoder and prediction network initialization. Two transfer schemes are shown in this figure: a) hi-IN CE + hi-IN LM initialization where the encoder is initialized with hi-IN CE model b) en-US CE + hi-IN LM initialization where the encoder is initialized with en-US CE model.}
\label{Enc_LM_init}
\end{figure}

\subsection{Encoder and prediction network initialization}
In the previously described transfer learning methods, only the encoder is initialized with a pretrained model. In this section, we will discuss initializing both encoder and prediction network with pretrained models as shown in Fig. \ref{Enc_LM_init}. 

The prediction network is initialized with a pretrained LSTM LM which is trained on an external text corpus %(obtained from different external sources) 
as a language model using grapheme units, referred to as hi-IN LM. The sentence count (number of repetitions of the sentence or the query) differ significantly from one source to the other. In order to avoid biasing the LM towards the source with large sentence counts, we only select unique sentences for LM training. After selecting the unique sentences, the hi-IN LSTM LM is trained on approximately $200$ million words.

The encoder is initialized with either hi-IN or en-US CE model as shown in Fig.\ref{Enc_LM_init}. For the purpose of compact representation, hi-IN and en-US CE model training is shown in a single block in Fig. \ref{Enc_LM_init}. They do not share any parameters and are trained independently. The hi-IN CE model is trained with hi-IN acoustic data and grapheme targets, and en-US CE model is trained with en-US acoustic data and word piece targets. The  configuration of initializing the prediction network with hi-IN LM and encoder with hi-IN CE model is referred to as hi-IN CE + hi-IN LM initialization. Similarly, the configuration of initializing the prediction network with hi-IN LM and the encoder with en-US CE model is referred to as en-US CE + hi-IN LM initialization.

We did not explore initializing the prediction network with en-US LSTM LM, as en-US and hi-IN lexical units differ (grapheme vs word piece), resulting in the input embedding matrices being significantly different, and thereby initializing the prediction network with en-US LSTM LM might not be beneficial. We also did not experiment initializing encoder with en-US RNN-T model in the context of encoder and prediction network initialization, as our experiments suggested that en-US CE initialization is better than en-US RNN-T initialization as discussed later in Section \ref{discussion_res}.

\section{Experimental setup}
\label{expt_setup}
The hi-IN models are trained with approximately $4$ million utterances amounting to few thousand hours of speech data. The speech data is distorted by noise to achieve robustness to noisy conditions. The en-US models were trained with $65000$ hours of data. The hi-IN test set contains $17619$ utterances consisting of five different scenarios including phrasal, conversational and code-mixed utterances. Training and test utterances are anonymized to remove any personally identifiable information.

We use $80$-dimensional log Mel filter bank features computed every $10$ milliseconds (ms). Eight vectors are stacked together to form $640$-dimensional acoustic features fed to the encoder. The frames are shifted by $30$ms. The hyper-parameters such as number of layers, layer dimension, frame size was tuned for en-US model \cite{hu_ce_init} and we adopted the same parameters for Hindi. All encoders have six LSTM layers with $1600$ hidden dimension and  $800$ projection dimension. All prediction networks have two LSTM layers with same cell dimension as encoders. Such a model setup follows the en-US work in \cite{Li2020Developing}. All models are evaluated after training for $6$ sweeps of the training data. 

The hi-IN grapheme targets contain all the unique graphemes in the Hindi native script. We also include grapheme targets with $B\_$ prefix to be able to segment the grapheme sequence into word sequence. Some research works use {\textless space\textgreater} symbol, however, we observed better accuracy with $B\_$ prefix based grapheme targets. A total of $130$ grapheme targets are obtained by combining the the original Hindi graphemes, graphemes with $B\_$ prefix and {\textless blank\textgreater} symbol. The word piece targets for en-US model is obtained by using byte pair encoding \cite{bpe} algorithm as described in \cite{hu_ce_init}.

We also report the word error rate (WER) on hybrid ASR model trained with same amount of data. The AM consists of $6$ layers of latency-controlled bidirectional LSTM \cite{LC_BLSTM} with $1024$ hidden dimension and $512$ projection dimension. AM is CE trained followed by EMBR training. The softmax layer has $9212$ senone labels. $80$-dimensional log Mel filter bank features are computed every $10$ms. Frame skipping \cite{Miao16} is done by a factor of $2$. Run-time decoding is performed using a 5-gram language model. 

\begin{table}
\begin{tabular}{|c|c|}
\hline 
Experiment & WER\tabularnewline
\hline 
Random initialization & 26.53\tabularnewline
\hline 
en-US RNN-T initialization & 22.97\tabularnewline
\hline 
en-US CE initialization & 22.38\tabularnewline
\hline 
Two-stage initialization with senone targets & 22.31 \tabularnewline
\hline 
Two-stage initialization with grapheme targets & \textbf{21.89} \tabularnewline
\hline 
hi-IN CE + hi-IN LM initialization & 24.29\tabularnewline
\hline 
en-US CE + hi-IN LM initialization & 22.63\tabularnewline
\hline 
Hybrid model & 22.32\tabularnewline
\hline 
\end{tabular}\caption{WER [\%] on hi-IN test sets for random initialization, different transfer
learning methods and the hybrid model.}
\label{full_results}
\vspace{-0.6cm}
\end{table}

\section{Discussion of results}
\label{discussion_res}
Table. \ref{full_results} shows WER for different transfer learning methods on hi-IN test sets. en-US CE initialization outperforms random initialization with $15.6\%$ relative WER (WERR) reduction. en-US RNN-T initialization is better than random, while is slightly inferior to en-US CE initialization. This could be because the en-US RNN-T encoder representations are influenced by en-US prediction network representations, as they are trained jointly. However, en-US CE model is trained in isolation and could serve as a better initialization model for the encoder. Two-stage transfer learning  with grapheme targets performs better than the rest of the methods with $17.4\%$ WERR reduction over random initialization. The pretraining method, hi-IN CE + hi-IN LM initialization improves over random initialization showing the importance of pretraining. The en-US CE + hi-IN LM initialization is better than pretraining, however, is inferior compared to en-US CE initialization, contrary to our expectation. The reason for such a behaviour is not known and we look to investigate this further in the future. With the improvements obtained from transfer learning, the WER of hi-IN RNN-T model (with transfer learning) is in parity with the hybrid model.

\subsection{Efficacy of TL for different amount of training data}
To study the efficacy of the TL with different amount of training data, we sample the original hi-IN data into smaller data-sets consisting of $50$ hours, $500$ hours and $1000$ hours. The RNN-T models are trained with random initialization and en-US CE initialization. The corresponding WERs on hi-IN test-set are shown in Table. \ref{tl_small}. en-US CE initialization shows $42.7\%$ WERR reduction for $50$ hour training over random initialization. The large WER gains with smaller training sets could be due to better RNN-T training convergence with TL as discussed in the next section. The above results also show a need for larger training data in RNN-T models. 

\begin{table}
\begin{tabular}{|c|c|c|c|}
\hline 
 & 50 hours & 500 hours & 1000 hours\tabularnewline
\hline 
\hline 
Random initialization & 83.77 & 69.32 & 51.56\tabularnewline
\hline 
en-US CE initialization & 47.96 & 35.07 & 32.75\tabularnewline
\hline 
\end{tabular}\caption{WER [\%] comparison on hi-IN test set between random and en-US CE initialized
RNN-T models trained with 50 hours, 500 hours and 1000 hours.}
\label{tl_small}
\vspace{-0.8cm}
\end{table}

\begin{figure}
%\begin{center}
%\hspace{-0.5 cm}
\includegraphics[scale=0.47]{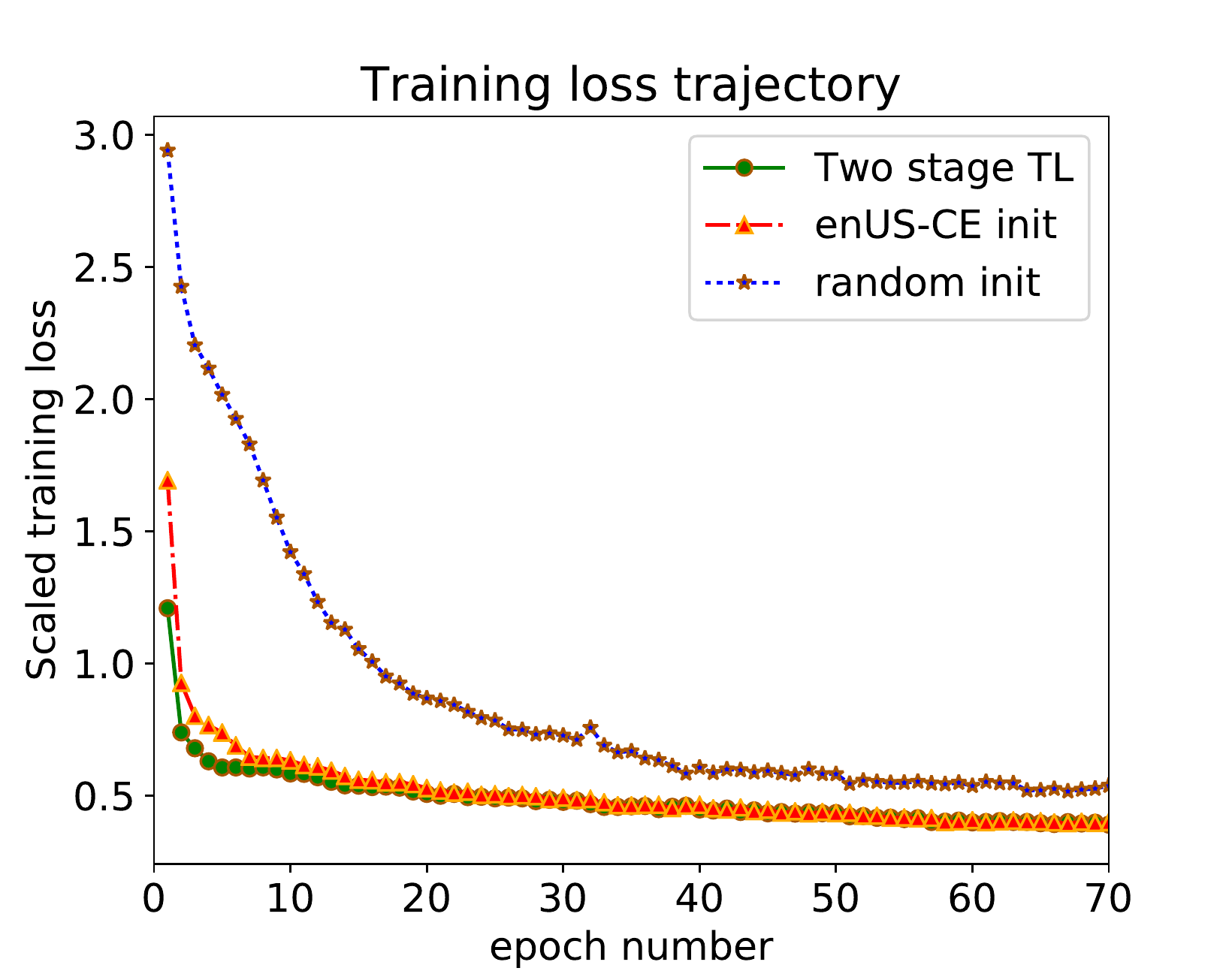}
%\end{center}
\caption{Training loss with increasing epoch number for random initialization, en-US CE initialization and Two stage TL  with grapheme targets.}
\label{train_loss}
\vspace{-0.4cm}
\end{figure}

\subsection{RNN-T training convergence}
\label{rnnt_training}
The training loss with increasing epochs for random initialization, en-US CE initialization and Two-stage TL with grapheme targets is shown in Fig. \ref{train_loss}. Each epoch is trained with $60$ hours of non-overlapping speech data. The training parameters such as learning rate, mini-batch size are identical for all three methods shown in Fig. \ref{train_loss}.  The training loss converges much faster with the transfer learning methods than random initialized models. The Two-stage TL converges  faster than en-US CE initialization. It is also interesting to note that the first epoch training loss is much lower for Two-stage TL than others, thereby indicating the superiority of initialization models in Two-stage TL.

\section{Conclusion}
\label{conclusion}
In this paper, we explore transfer learning methods for RNN-T models. Our motivation is to leverage well-trained en-US models to bootstrap hi-IN RNN-T models and also to stabilize the hi-IN RNN-T model training. We evaluated the following transfer learning methods: a) en-US CE initialization b) en-US RNN-T initialization c) Two-stage transfer learning and d) Encoder and prediction network initialization.  Based on the WER gains and training convergence, we propose Two-stage learning approach with grapheme targets as the preferred transfer learning strategy. The experiments on smaller data-sets and training loss convergence reveal the importance of transfer learning for low-resource RNN-T models. The methods discussed in this paper can be generalized to other low-resource languages as well. In future, we plan to explore other transfer learning methods and its extension to multi-lingual RNN-T models.
\bibliographystyle{IEEEtran}

\bibliography{my.bib}

\end{document}